\documentclass[useAMS,usenatbib]{eas}
\usepackage{graphicx}
\usepackage{aas_macros}  
\usepackage{natbib}
\usepackage[dvipsnames]{color}
\usepackage{amssymb,amsmath}

\begin{document}

\title{Spectroscopic ages and metallicities of galaxies} 
\runningtitle{Koleva \etal: Spectroscopic ages and metallicities}
\author{M. Koleva}\address{IAC, La Laguna and
      Dept. de  Astrof\'{\i}sica, Univ.  de  La Laguna,
      E-38200,   Tenerife,  Spain }
\secondaddress{Universit\'e Lyon~1, Villeurbanne; CRAL, Observatoire de Lyon, France ; 
  CNRS, UMR 5574 }
\author{A.\,G.\,Bedregal}\address{Dept. de Astrof\'{\i}sica y CC. de la Atm\'{o}sfera, 
  Univ. Complutense de Madrid, E-28040, Spain }
\author{Ph. Prugniel}\sameaddress{2}
\author{S. De~Rijcke}\address{Dept. of Physics \& Astronomy, Ghent University, Krijgslaan 281, S9, B-9000, Belgium}
\author{W. W. Zeilinger}\address{Institut
  f\"{u}r Astronomie, Universit\"{a}t Wien, T\"{u}rkenschanzstrasse
  17, A-1180 Wien, Austria }
\begin{abstract}
Dwarf galaxies are generally faint. To derive their age and
metallicity distributions, it is critical to optimize the use of any
collected photon.  Koleva \etal, using full spectrum fitting, have
found strong  population gradients in some dwarf elliptical galaxies.
Here, we show that the population profiles derived with this
method are consistent and more precise with those obtained with
spectrophotometric indices. This allows studying fainter objects
in less telescope time.
\end{abstract}
\maketitle
\section{Introduction}

Stellar populations gradients of galaxies, not resolved into 
stars, are obtained by analyses of spatially resolved spectra (e.g. long-slit).
They have to be carried to the effective radius, at least,
where the signal becomes low. Koleva \etal\ \cite{koleva2009c}
investigated the effects of an inaccurate sky subtraction 
and data processing on the gradients and 
granted confidence in the results of Koleva \etal\ \cite{koleva2009b}.
To complete this validation, we compare analyses
made with two different methods: (1)
a well established, grid inversion of spectrophotometric
indices (SI); (2) full spectrum fitting (FSF).
Such a comparison was made earlier by 
Michielsen \etal\ \cite{michielsen2007} for some high quality central extractions. The present test probes the influence of the signal-to-noise ratio (S/N).

\section{Full spectrum vs. Lick indices fitting}

We analyse long-slit data obtained for nine S0 galaxies with FORS2/VLT. 
Some of them are dS0s. The data and their reduction are presented 
in Bedregal \etal\ \cite{bedregal2006}.
Bedregal \etal\ \cite{bedregal2011} obtained ages and metallicities profiles using
$H_\beta$ and MgFe$^\prime$ Lick indices. Here, we analyse the same data with
the FSF package ULySS \citep[Koleva \etal\ ][]{koleva2009a}.
The two sets of profiles are 
presented on Fig.1 for four objects. The agreement 
is excellent for all the galaxies.
The  FSF is a more precise method, because it uses
all the information in the spectrum, works with a higher 
total S/N. It allows to carry the analysis to larger galactocentric distances.
Discussions about the actual results can be found in
Bedregal \etal\ \cite{bedregal2011} and Koleva \etal\ \cite{koleva2011}. \\

\begin{figure}
\label{fig:1}
\includegraphics[width=0.24\textwidth]{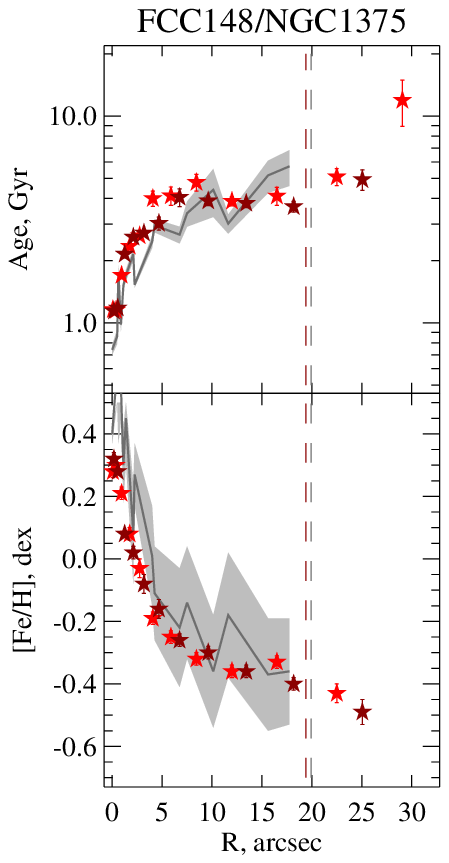}
\includegraphics[width=0.24\textwidth]{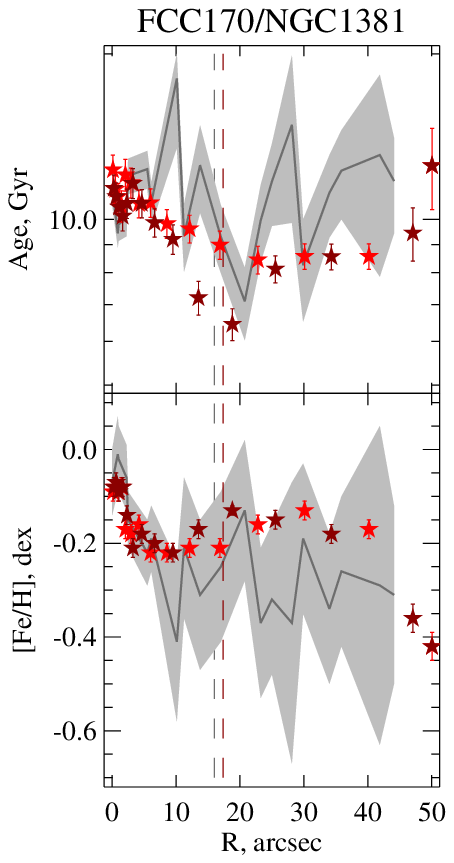}
\includegraphics[width=0.24\textwidth]{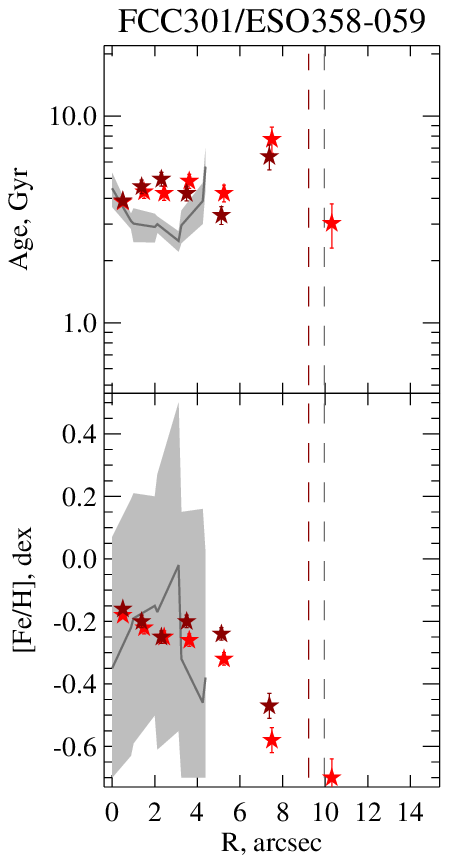}
\includegraphics[width=0.24\textwidth]{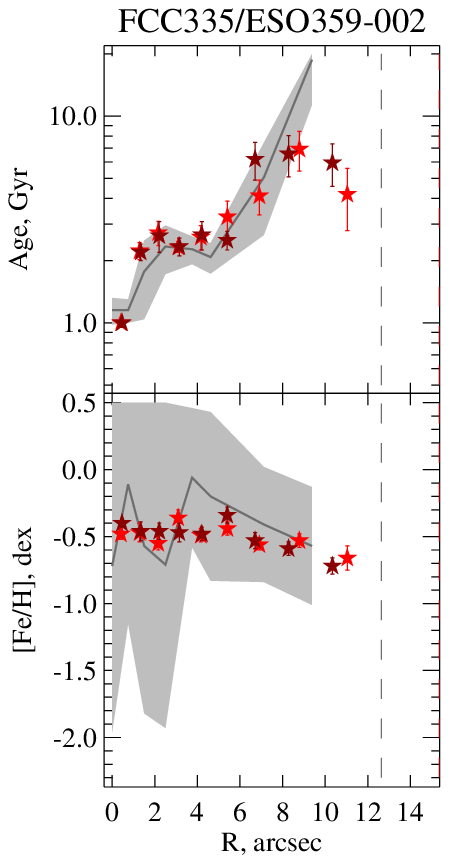}
\caption{Population radial profiles derived with SI (Bedregal \etal\ \cite{bedregal2011}, grey continuous lines, the shaded area marks the errors) and ULySS
(red stars). The top/bottom panels show ages/metallicities along the radii.
The effective radii are plotted with dashed lines.  }
\end{figure}

\small{
MK and AGB have been supported by the Programa 
Nacional de Astronom\'{\i}a y Astrof\'{\i}sica of the Spanish Ministry 
of Science and Innovation under grant \emph{AYA2007-67752-C03-01/03}}.

\end{document}